\documentclass[aps,prb,twocolumn,showpacs]{revtex4}

\usepackage{graphicx}
\usepackage{amssymb}
\usepackage{bm}

\newcommand{\smeq}{\! = \!}
\newcommand{\smneq}{\! \neq \!}
\newcommand{\smpl}{\! + \!}
\newcommand{\smmi}{\! - \!}

\newcommand{\ve}{\varepsilon}

\newcommand{\Ef}{E_{\mathrm{F}}}

\newcommand{\be}{\begin{equation}}
\newcommand{\ee}{\end{equation}}
\newcommand{\bea}{\begin{eqnarray}}
\newcommand{\eea}{\end{eqnarray}}
\newcommand{\Ha}{{\hat H}}

\newcommand{\ci}{\mathrm{i}}

\begin{document}

\title{Spin Hall effect in clean two dimensional electron gases with Rashba spin-orbit coupling}
\author{A. Reynoso}
\affiliation{Instituto Balseiro and Centro At\'omico Bariloche,
Comisi\'on Nacional de Energ\'{\i}a At\'omica, 8400 San Carlos de
Bariloche, Argentina.}
\author{Gonzalo Usaj}
\affiliation{Instituto Balseiro and Centro At\'omico Bariloche,
Comisi\'on Nacional de Energ\'{\i}a At\'omica, 8400 San Carlos de
Bariloche, Argentina.}
\author{C. A. Balseiro}
\affiliation{Instituto Balseiro and Centro At\'omico Bariloche,
Comisi\'on Nacional de Energ\'{\i}a At\'omica, 8400 San Carlos de
Bariloche, Argentina.}
\date{submitted 30 November 2005}

\begin{abstract}
We study the spin polarization induced by a current flow in clean
two dimensional electron gases with Rashba spin-orbit coupling.
This \textit{geometric} effect originates from special properties
of the electron's scattering at the edges of the sample. In wide
samples, the spin polarization has it largest value at low
energies (close to the bottom of the band) and goes to zero at
higher energies. In this case, the spin polarization is dominated
by the presence of evanescent modes which have an explicit spin
component outside the plane. In quantum wires, on the other hand,
the spin polarization is dominated by interference effects induced by multiple
scattering at the edges. Here, the spin polarization is quite
sensitive to the value of the Fermi energy, especially close to the point where a new
channel opens up. We analyzed different geometries and
found that the spin polarization can be strongly enhanced.
\end{abstract}

\maketitle

\section{Introduction}
The possibility to create and manipulate spin currents or to
induce spin polarization at the edges of a semiconductor by simply
applying an external charge current has created great expectation
in the field of spintronics---a fast developing area aimed to
build up new technologies based on the
 manipulation of the electron's
spin.\cite{Spintronicsbook} The phenomenon of current induced spin
polarization (CISP)---also called  spin Hall effect
(SHE)  when referred to the perpendicular
polarization in two dimensional systems---was predicted some time ago by D'yakonov and
Perel.\cite{DyakonovP71} However, recent theoretical
predictions\cite{Hirsch99,MurakamiNZ03,SinovaCSJM04} and, more
lately, the experimental observation of the
effect\cite{KatoMGA04,WunderlichKSJ05,KatoMGA05-lshaped,SihMKLGA05} have renewed the
interest on the subject and triggered a tremendous among of
work during the last two
years.$^{10-38}$
Although some of the initial controversies has been settled, some aspects of the
problem are not completely understood yet.

Current induced spin polarization may occur in systems with strong
spin-orbit (SO) coupling,\cite{Edelstein90} where the electron's momentum and its spin
are coupled.\cite{Winkler_book} The simplest setup for observing a CISP consists of a
 bar-shaped sample connected to two terminals. When a charge current
flows through the system, a spin polarization develops at the
lateral edges of the sample with different signs on each side. The mechanisms
leading to this polarization can be of different nature. In the context of the SHE, they
have been classified as extrinsic or intrinsic. The former occurs in dirty
semiconductors, when the presence of impurities generates a spin
dependent scattering which deflects different spin projections in
different directions.\cite{DyakonovP71,Hirsch99}
The latter, on the other hand, is due to the presence of an intrinsic SO coupling,
which is non-local in space and results from the breaking of the inversion
symmetry of the sample or, in the case of an heterostructure, of the confining
potential.\cite{Winkler_book} This is the mechanism
underlying the prediction of dissipansionless spin currents in p-doped
semiconductors\cite{MurakamiNZ03} and in two-dimensional electron
gases (2DEG) with Rashba spin-orbit coupling.\cite{SinovaCSJM04}

In the last case, it was argued that the spin Hall
conductivity, defined as the ratio between the transverse spin current (generated by the
charge current flow) and the external electric field, have the
universal value $e/8\pi$.\cite{SinovaCSJM04} Both, the fact of this value being
independent of the scattering time, and the fact that the spin current is not conserved
in systems with SO, lead to some controversy.
Furthermore, because of the 
non-conservation of the spin current, the very existence of CISP
is not obvious. Later works, however, have shown that the spin Hall
conductivity, as obtained in \textit{infinite} homogeneous systems, is very sensitive to
disorder and that, in the case of Rashba coupling, it vanishes even in the weak disorder
limit.\cite{InoueBM04,MishchenkoSH04,Khaetskii04,Rashba04-2,ChalaevL05,
BernevigZ05,RaimondiS05,MalshukovC05} It is important to emphasize at this point
that the finite size
of the sample was not taken into account explicitly in early work but have been
included in more recent
ones.\cite{NomuraWSKMJ05,NikolicSZS05,NikolicZS05,ShengST05,NikolicZW05,AdagideliB05,
NikolicZS05_imaging,TseFZS05,LozanoS05}

Recent experiments, on the other hand, succeed in observing CISP in
semiconducting heterostructures\cite{WunderlichKSJ05,SihMKLGA05} and in GaAs thin
films.\cite{KatoMGA04} The optically detected magnetization at the edge of a finite
sample  clearly shows  the existence of the effect. In addition,  CISP at
the corner of a L-shaped sample was also observed,\cite{KatoMGA05-lshaped}
showing the effect is not restricted to a straight sample configuration. The nature
of the observed SHE is, nevertheless, still unclear. Arguments in favor of an
intrinsic SHE have been put forward in Refs.
[\onlinecite{WunderlichKSJ05,NomuraWSKMJ05}] while others
have argued in favor of an extrinsic explanation,
Refs. [\onlinecite{KatoMGA04,SihMKLGA05,EngelHR05}]

The debate on the physical origin of the CISP triggered a number of numerical
studies.\cite{HankiewiczMJS04,NikolicSZS05,UsajB05_SHE,NikolicZS05,
ShengST05,NikolicZW05,AdagideliB05,NomuraWSKMJ05,NomuraSJNM05,NikolicZS05_imaging,
TseFZS05}
For the case of clean 2DEG with Rashba SO-coupling, it was shown in Ref.
[\onlinecite{UsajB05_SHE}] that spin polarization near the sample edge can develop
\textit{kinematically}: a ballistic current that unbalance the occupation of
positive and negative velocity states leads to a net polarization perpendicular to the
2DEG\cite{note0}---similar results were obtained in Ref.
[\onlinecite{NikolicSZS05}] for the case of narrow samples.
This effect is due to the non-trivial structure of the wave functions as result of the boundary condition.
It is important to emphasize that in this clean system, the CISP does \textit{not}
result from the presence of a `bulk' spin current.
Therefore, when interpreting the CISP in ballistic system as resulting from the action
of a `force', it is important to make clear that the boundary condition plays a critical
role. \cite{NikolicZW05,Shen05} In this sense, it is worth pointing out that in the case
of a bulk system, a particle with spin `up' in the $\hat{z}$-direction, will not drift
but rather follow an oscillatory trajectory.\cite{SchliemannLW05}

Within this context, it is then important to clarify what is the linear response of  a
ballistic system, what is the sample size dependence of the
CISP or how the geometry can generate different profiles for the spin
polarization. Here, we address these questions by studying the spin response of
nanoscopic systems to an external bias using the Keldysh
formalism.\cite{Jauho_Book,Pastawski92}

The paper is organized as follows: the main features of the electron's scattering at
the surface is revisited in Section II for the  case of semi-infinite 2DEGs.
Finite size samples and geometry effects are analyzed
in Section III. We summarize and conclude in Section IV.

\section{Spin polarization in wide systems}
\subsection{Semi-infinite systems}
Our starting point is a system with the simplest geometry
able to show CISP: a semi-infinite 2DEG. Since we consider the ballistic regime,
the reflection at the sample's edge is the only scattering process. As we
discuss below, this scattering process has very unusual
properties.\cite{KhodasSF04,RamagliaBCFP04,ChenHPDG05,UsajB05_SHE}
The most relevant are: (i) the appearance of evanescent modes
localized at the edge and (ii) the mixing of bulk states from different bands. As we
shall see, this \textit{geometric} effect is enough to lead to spin
polarization in the presence of an external charge current.

\begin{figure}[t]
   \centering
   \includegraphics[width=.35\textwidth]{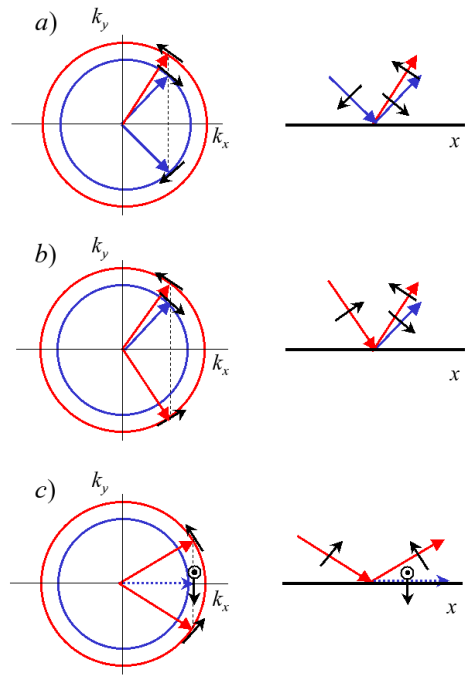}
   \caption{(color online) Schematic representation of the scattering process at the
edge. Panels a) an b) correspond to incident waves with $k_x\!<\!k_{+}$. In those cases,
two waves are reflected, one on each band. Panel c) shows the case of $k_x>k_{+}$. Here,
there is only one reflected plane wave but an evanescent mode localized at the edge
appears. The spin of the localized mode has a nonzero component both in the $y$ and the
$z$ direction.}
   \label{fig1}
\end{figure}

Let us then consider a 2DEG with Rashba SO-coupling described by
the following Hamiltonian 
\be 
\Ha\smeq\frac{p_{x}^{2}\smpl
p_{y}^{2}}{2m^{*}}\smpl V(\bm{r})\smpl \Ha_{\mathrm{SO}}
\label{Ht} 
\ee 
where $m^{*}$ is the effective mass and $V({\bf
r})$ is the lateral confining potential. The last term accounts
for the SO interaction,\cite{Rashba60,BychkovR84}
\be
\Ha_{\mathrm{SO}}\smeq\frac{\alpha}{\hbar}(p_{y}{\hat
\sigma}_{x}\smmi p_{x}{\hat \sigma}_ {y})
 \ee
 with $\alpha $ the
Rashba coupling constant. It is readily seen from the above
equations that the SO coupling acts as a strong in-plane magnetic
field, proportional to the momentum an perpendicular to it,
$\bm{B}_\mathrm{int}\smeq(g\mu_B/2)^{-1}\alpha/\hbar
(p_{y},-p_{x})$. The presence of this effective field breaks the
spin degeneracy and leads to two non-degenerate conduction bands.
In an homogeneous 2DEG ($V(\bm{r})\smeq0$) the eigenstates are
given by \be \Psi_{\bm{k},\pm
}(\bm{r})\smeq\frac{1}{\sqrt{2A}}{\rm e}^{{\rm i}\bm{k}\,\cdot
\bm{r}}\left( {{\pm {\rm e}^{-{\ci}\phi /2} \atop {\rm
e}^{{\ci}\phi /2}}}\right) 
\ee 
with ${\rm e}^{{\ci}\phi}\smeq(k_{y}\smmi{\ci}k_{x})/k$ and $A$ the area of the system.
Note that the electron's spin points in the direction of
$\bm{B}_\mathrm{int}$, \textit{i.e.} it is perpendicular to the
momentum (see Fig. \ref{fig1}). The corresponding eigenvalues are
\be \varepsilon _{\pm }({\bf k})\smeq\hbar ^{2}k^{2}/2m^{*}\!\pm
\!\alpha k\,. \ee Then, for a given energy $\ve$, there are two
characteristic wavevectors, $k_{+}$ and $k_{-}$ (with
$k_{+}\!<\!k_{-}$). We now introduce the sample's edge by the
following potential \be V({\bf r})\smeq\left\{
\begin{array}{cc}
0&\mathrm{for}\,\, y>0 \\
\infty &\mathrm{for}\,\, y<0
\end{array}
\right.\,. \ee

 Because of the translational invariance of the
system along the $x$-axis, an incident wave $\Psi _{(k_x,k_y),\eta
}({\bm r})$, where $\eta\smeq\pm$ is the band index, is reflected
conserving the $x$-component of the momentum, $k_{x}$. The
continuity of the wavefunction at the edge requires $\psi _{\bm{
k},\eta}(x,0)\smeq0$, where $\psi _{\bm{ k},\eta}(\bm{r})$ is the
total wavefunction. Since two states on the same band with the
same $k_x$ have different spin orientations, the boundary
condition can only be satisfied if the reflected wave is a linear
combination of the two modes $\Psi _{(k_x,-k_y),\eta}(\bm{r})$ and
$\Psi_{(k_x,-k'_y),-\eta}(\bm{r})$ as schematically shown in Figs.
\ref{fig1}a and \ref{fig1}b. This superposition of plane waves
with different spin generates oscillations of the spin density.
This interference effect have interesting properties that
manifest more clearly in narrow wires (see next Section).
\begin{figure}[t]
   \centering
   \includegraphics[width=0.45\textwidth,angle=-90,clip]{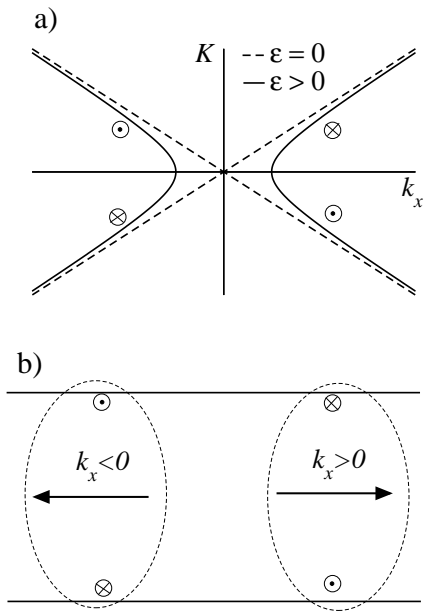}
   \caption{a) Constant energy surface in the $K-k_x$ plane for $\ve\smeq0$ and $\ve\smneq0$. $\odot$ and $\otimes$ indicates spin `up' and spin `down' in the $z$-axis, respectively. b) Schematics of the eigenstates of a wide sample for $k_x\!>\!k_+$. Notice that each states has opposite sign of the spin on opposite sides of the sample and that the sign of the spin depends on the sign of $k_x$.}
   \label{fig2}
\end{figure}
For the geometry we are considering here (or for wide samples), it is more
convenient to analyze first the case of large incident angles.
It is clear from Fig.\ref{fig1}c that for $k_x>k_{+}\smeq[(2\hbar
^{2}\varepsilon /m^{*}\smpl\alpha ^{2})^{\frac{1}{2}}\smmi|\alpha|]m^{*}/\hbar ^{2}$, there are no propagating modes in the
`$+$' band. However, for those values of $k_x$, the Sch\"odringer equation admittes an evanescent wave solution given by
\be
\Psi_{\mathrm{ev}}(\bm{r})\smeq\frac{1}{C}\,{\rm e}^{\ci\,k_{x}x} {\rm e}^{K y}\left( {{a \atop b}}\right),
\ee
where
\be
K\smeq\pm\sqrt{k_x^2\smmi k_{+}^2}\,,\qquad |\alpha|k_{+}a\smeq\ci\alpha (k_x\smmi K)b\,,
\ee
$C$ is a normalization constant and $ |a|^2\smpl|b|^2\smeq1$.
A generic energy surface in the $K$-$k_x$ space is shown in Fig.\ref{fig2}.
While in bulk this solution does not have any physical meaning, in the presence of a
surface it \textit{must} be included in order to satisfy the boundary condition. The
total wavefunction is then a linear superposition of $\Psi _{(k_x,k_y),-}(\bm{r})$,
$\Psi_{(k_x,-k_y),-}(\bm{r})$ and $\Psi_{k_x,\mathrm{ev}}(\bm{r})$.
The evanescent modes have two interesting properties which are key ingredients in
leading to CISP in wide samples---the solutions for smaller incident angles share some
of those properties, as we discuss below, but are less obvious.
First, we notice that the evanescent modes have an explicit spin component in
the $z$-direction. This is clear from the fact that
\be
\frac{|a|^2}{|b|^2}\smeq\frac{k_x\smmi K}{k_x\smpl K}\smneq1\,.
\ee
\textit{Moreover, the sign of the spin projection depends only on the sign of $k_x$ and
$K$}. Since the sign of $K$ is fixed by the condition that the wavefunction
remains finite for $y\rightarrow\infty$ ($K\!<\!0$ in our case), the
sign of the spin projection is giving only by the sign of $k_x$. Then, electrons with
$k_x\!>\!0$ have spin `up' ($a^2/b^2\!>\!1$) and those with $k_x\!<\!0$ have spin `down'
($a^2/b^2\!<\!1$). Note that for $\varepsilon=0$ the evanescent states are fully
polarized, $a\smeq1$ for $k_x\!>\!0$ and $b\smeq1$ for $k_x\!<\!0$ and that they are
the only states available since $k_{+}\smeq0$.
It is worth mentioning that in the case of a wide but finite sample, the solution
with $K\!<\!0$ dominates at the lower edge while the one with $K\!>\!0$ does it at the
upper edge. Then, for a given state (value of $k_x$) the spin projection have opposite
signs in opposites sides of the sample. This is schematically shown in Fig. \ref{fig2}b.
Notably, the above results do not depend on the sign of $\alpha$. This result is
consistent with the fact that the effective `force'\cite{NikolicZW05,Shen05} acting on a
wave package is proportional to $\alpha^{2}$.

\begin{figure}[t]
   \centering
   \includegraphics[width=.4\textwidth]{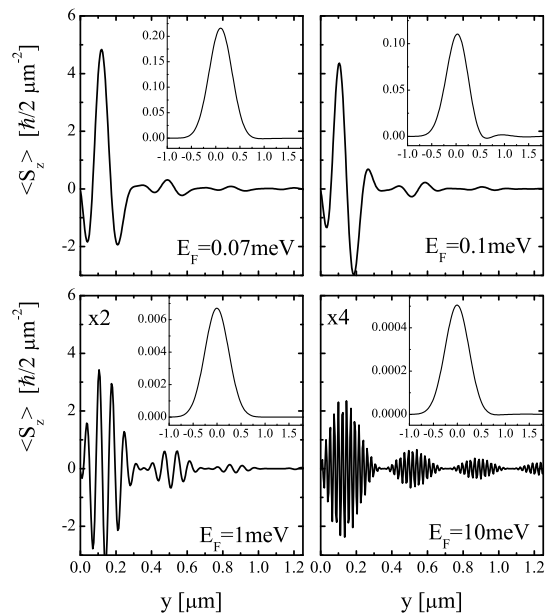}
   \caption{Current induced spin polarization in a semi-infinite 2DEG for different
values of $\Ef$. We used $\alpha\smeq10$meVnm, $m^{*}\smeq0.068m_0$ and
$eV\smeq0.13$meV. The inset correspond to a convolution with a Gaussian of
$\mathrm{rms}=0.25\mu$m. The CISP shown in the bottom panels were multiplied by the
factors indicated in the figure for the propose of comparison.}
   \label{fig3}
\end{figure}
\begin{figure}[t]
   \centering
   \includegraphics[width=.4\textwidth]{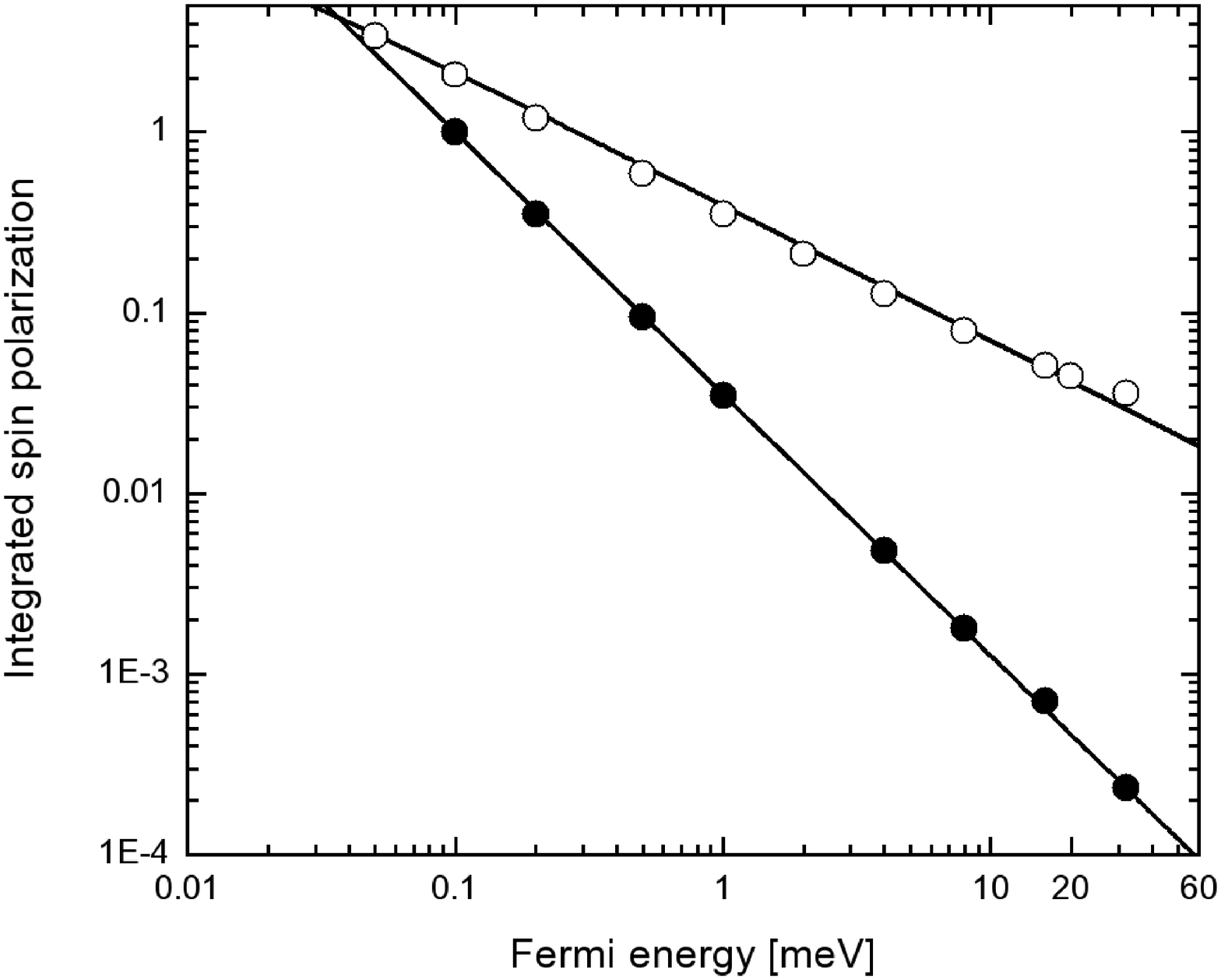}
   \caption{Integrated spin polarization $\mathcal{I}$ as a function of
$\Ef$ for $\alpha\smeq10$meVnm. The solid dots show the total spin polarization while
the open dots show the contribution of the evanescent modes. Notice $\mathcal{I}$
decays with $\Ef$ as a power law (the lines are a guide to the eyes).}
   \label{integrada}
\end{figure}

This simple analysis of the evanescent modes shows that there is an \textit{intrinsic
asymmetry} that arises from the combined effect of the breaking of translational
invariance and the intrinsic chirality introduced by the spin-orbit interaction,
$V_{\mathrm{SO}}\propto({\vec p}\times{\vec \sigma})\cdot{\hat z}$. This is \textit{all}
what is
needed to justify the appearance of CISP at $\varepsilon\smeq0$. For higher energies,
the states without evanescent modes must be included. Although each of those states do
not have a well defined spin component in the $z$-direction (it oscillates as a
function of $y$), the sum of all the states with $k_x\!<\!k_{+}$ leads to a non-zero
contribution which has the opposite sign that the one corresponding to the
evanescent
modes (see Fig \ref{integrada}).
Despite it is hard to prove analytically that such contribution  is non-zero, it is
rather straightforward to show that each state has the property that the
$z$-component of the spin density changes sign when $k_x$ does it. 

In order to calculate the CISP in this ideal geometry, we consider a biased 2DEG where
the voltage drops at the contacts (not included in the calculation) so the electric
field is \textit{zero} inside the sample. Then, in the energy interval [$\Ef\smmi
eV/2,\Ef\smpl eV/2$], the carriers injected from the left contact, which has the highest
chemical potential, occupied only the states with
$k_{x}\!\geq\!0$.\cite{note2,Dattabook} The CISP is then given by
\be
\langle {S}_{z}(y)\rangle \smeq \frac{\hbar}{2}
\sum_{\bm{k},\eta;k_{x}>0}\langle {\hat{\sigma}}_{z}\rangle_{\bm{k},\eta}\,
F_{\bm{k},\eta}(E_{\mathrm{F}},eV)
\label{spinacc}
\ee
where $\langle {\hat{\sigma}}_{z}\rangle_{\bm{k},\eta}\smeq\psi _{\bm{ k}%
,\eta}^{\dagger }(\bm{ r}){\hat{\sigma}}_{z}\psi _{\bm{ k},\eta}(\bm{ r}) $ with $F_{\bm{k},\eta}(\Ef,eV)\smeq\Theta
(\Ef\smpl eV/2\smmi\varepsilon_{\eta}(\bm{k}))\smmi\Theta (\Ef\smmi
eV/2\smmi\varepsilon_{\eta}(\bm{ k}))$.
In what follows we calculate $\langle S_{z}(y)\rangle$ in linear response by numerical
integration
of the Schr\"{o}dinger equation using finite differences. The advantage of this method
is twofold: a) it is not restricted to a square well confining potential; b) it can
be easily generalized to include the contacts (see next section).
All quantities presented below are obtained from the one particle propagators
 which are calculated using a continues fraction method (see
Ref.[\onlinecite{UsajB04_focusing}] for details).

Figure \ref{fig3} shows $\langle S_{z}(y)\rangle$ for different values of $\Ef$ and
for $\alpha\smeq10$meVnm.
The insets correspond to the same results convoluted with a Gaussian. It is clear from
the figures that the contribution from the evanescent modes is large at low energies
and that the CISP goes to zero when the energy is increased.
The decay of the CISP with energy is shown in Fig. \ref{integrada} where we plotted the
integrated spin polarization, defined as
\be
\mathcal{I}\smeq\int_{0}^{\infty}\langle {S}_{z}(y)\rangle\, \mathrm{d}y
\ee
as a function of $\Ef$. Both, the total integrated spin
polarization ($\mathcal{I}_\mathrm{T}$, solid
dots) and the one coming from the evanescent modes ($\mathcal{I}_\mathrm{ev}$, open
dots) are shown. Clearly, the decay is well described by a power law. The fact that
$\mathcal{I}_\mathrm{T}$ is smaller than $\mathcal{I}_\mathrm{ev}$ indicates that the
spin polarization of the evanescent modes is opposite in sign to the rest of the modes.
The decay of
$\mathcal{I}_\mathrm{ev}$ can be fitted very well as $\mathcal{I}_\mathrm{ev}\smeq
b(\Ef\smpl \alpha^{2}m^{*}/2\hbar)^{-\frac{3}{4}}$ where $b$ is a
normalization
constant.\cite{UsajB05_SHE}
\subsection{Wide wires}
\begin{figure}[t]
   \centering
   \includegraphics[width=.52\textwidth,clip]{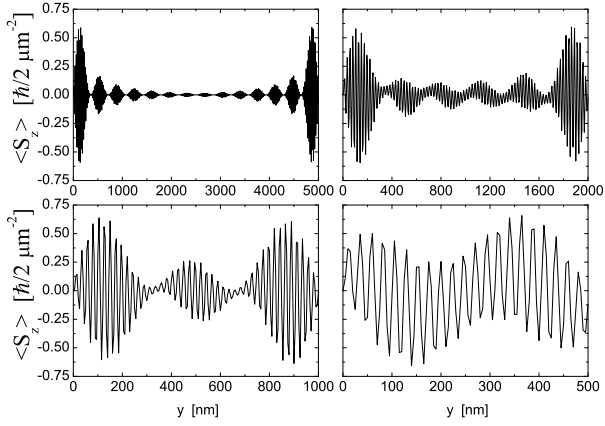}
   \caption{(color online) Current induced spin polarization for different sample's
width
$L_y\smeq5,2,1,0.5\mu$m and for $\alpha\smeq10$meVnm, $\Ef\smeq10$meV and
$eV\smeq0.13$meV. Notice the
CISP
is modulated on a length scale $L_{\mathrm{SO}}\smeq\pi\hbar ^{2}/m^{*}\alpha$.}
   \label{dependenciaconL}
\end{figure}

The case of a wide but finite sample can be analyzed in a very similar way.
The confinement potential is now given by
\be
V({\bf r})\smeq\left\{
\begin{array}{cc}
0&\mathrm{for}\,\, 0<y<L_{y}\\
\infty & \mathrm{otherwise}
\end{array}
\right.\,.
\ee
where $L_y$ is the width of the wire. The only difference here is that a
second boundary condition must be imposed, namely $\psi _{\bm{ k},\eta}(x,L_y)\smeq0$.
It
is clear from Fig. \ref{fig1} that for a given energy,  the four states having the same
$k_x$ are mixed. Here again  the evanescent modes must be
included when $k_x\!>\!k_{+}$ (in this case, both solutions with $K\!>\!0$ and $K\!<\!0$
are required).
Figure \ref{dependenciaconL} shows the CISP as a function
of the width of the sample. Notice the magnitude of the effect, when averaged over a
fixed area, is reduced as the sample
becomes wider and that the CISP is more localized at the sample's edge.  Besides the
oscillating pattern on the scale of the Fermi wavelength, there is a longer modulation
set by the SO coupling, $L_{\mathrm{SO}}\smeq\pi\hbar ^{2}/m^{*}\alpha$, which is
independent of
the system's size.

So far, we have shown that CISP is possible in clean systems with
SO coupling due to the special properties of the electron scattering at the sample's
edge. In very wide samples, the effect is essentially confined to a region of order
$L_{\mathrm{SO}}$ near the edges and its magnitude decay as the Fermi energy increases.
As the sample becomes narrower, however, multiple scattering at the edges leads to a
different profile for the CISP. It is important to emphasize that there is no
conceptual difference between both cases and that the picture
described so far is the basic physics underlying the mesoscopic SHE described in
Refs. [\onlinecite{NikolicSZS05,NikolicZS05_imaging,YaoY05,UsajB05_SHE}].

\section{Spin polarization in narrow samples}
Let us now discuss the CISP in finite samples. In
order to do that, we explicitly include the leads in our calculation. For simplicity, we
consider leads with the same microscopic parameters than the sample except for the SO
coupling parameter $%
\alpha $ which is taken to be zero inside the leads. Since the SO parameter $\alpha $
has a spacial variation at the sample-lead interfaces, which is
assumed to be only in the $x$-direction, the Hamiltonian includes a term
proportional to $-\ci\sigma _{y}\partial \alpha (x)/\partial
x$.\cite{note3} As mentioned before, to be able to describe
systems of arbitrary shape, and in order to
include the leads, it is convenient to reduce the continuum
effective mass model to a tight binding model
\begin{eqnarray}
\Ha &\smeq&\sum_{n\sigma }\varepsilon _{\sigma }c_{n\sigma }^{\dagger
}c_{n\sigma }^{}-\sum_{<n,m>\sigma }t_{nm}c_{n\sigma }^{\dagger }c_{m\sigma
}^{}+h.c.  \nonumber \\
&&-\sum_{n}\left\{ \lambda _{n,n+\widehat{y}}\left( {\rm i}\,c_{n\uparrow
}^{\dagger }c^{}_{(n+\widehat{y})\downarrow }\smpl{\rm i}\,c_{n\downarrow
}^{\dagger }c^{}_{(n+\widehat{y})\uparrow }\right) \right.  \nonumber \\
&&\left. -\lambda _{n,n+\widehat{x}}\left( c_{n\uparrow }^{\dagger }c^{}_{(n+%
\widehat{x})\downarrow }\smmi c_{n\downarrow }^{\dagger }c^{}_{(n+\widehat{x}%
)\uparrow }\right) \right\} +h.c.
\label{HTB}
\end{eqnarray}
where $c_{n\sigma }^{\dagger }$ creates an electron at site $n$ with spin $%
\sigma _{z}\smeq\sigma $ and energy $\varepsilon _{\sigma }\smeq4t$, $%
t\!_{n,m}\equiv t\smeq\hbar ^{2}/2m^{*}a_{0}^{2}$ for neighboring sites and $a_{0} $ is
the lattice parameter. The SO couplings are defined as follow:
$\lambda_{n,n+\widehat{x}}\smeq\lambda
\!_{n,n+\widehat{y}}\equiv \lambda \smeq\alpha /2a_{0}$ deep inside the sample, $%
\lambda_{n,n+\widehat{x}}\smeq(\alpha_n\smpl\alpha_{n,n+\widehat{x}})/4a_{0}$ and $\lambda_{n,n+\widehat{y}}\smeq\alpha_n/2a_{0}$ at the sample-lead interface and $\lambda \!_{n,m}\smeq0$ otherwise. The summation is carried out on a square
lattice where the coordinate of site $n$ is $\bm{ r}_{n}\smeq n_{x}\widehat{x}%
+n_{y}\widehat{y}$ with $\widehat{x}$ and $\widehat{y}$ the unit lattice
vectors in the $x$ and $y$ directions, respectively. Unless otherwise mentioned, we use
a smooth function to turn on the SO coupling $\alpha$ in order to avoid stron
g reflections at the sample-lead interfaces.
\begin{figure}[t]
   \centering
   \includegraphics[width=.45\textwidth]{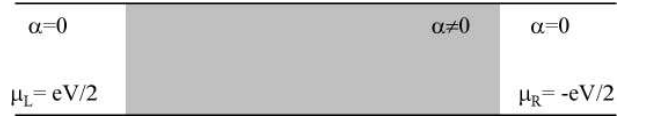}
   \caption{ Schematic representation of a finite quantum wire. The SO
coupling is non-zero in the dark area.}
   \label{scheme}
\end{figure}

When a voltage bias $V$ is applied between the leads, an electric current
flows through the sample. In this stationary but out of equilibrium state, the SO coupling
generates a pattern of local spin polarization. Within the Keldysh
formalism, the spin pattern is given by\cite{Mahan_Book,NikolicSZS05}
\begin{equation}
\langle S_\mathrm{a}(\bm{ r}_{n})\rangle \smeq\frac{-\ci
\hbar}{2}\mathrm{Tr}\left[ \sigma_\mathrm{a} \bm{G}_{nn}^{<}(t\smeq0)\right]
\label{Sgeneral}
\end{equation}
where the trace is taken on the spin variables, $\sigma_\mathrm{a} $ with $\mathrm{a}\smeq x,y,z$ are the
Pauli matrices and the lesser Green function $\bm{ G}_{nm}^{<}\left(
t\right) $ is a $2\!\times\! 2$ matrix with elements $\ci\langle c_{n\sigma
}^{\dagger }(t)c_{m\sigma'}^{}(0)\rangle $. In what follow
we consider small bias voltages and linearize Eq. (\ref{Sgeneral}). The Fourier
transform $\bm{ G}_{n,m}^{<}\left( \varepsilon \right) $ of the lesser Green
function is given by\cite{Jauho_Book,Pastawski92}
\begin{eqnarray}
\bm{ G}_{nm}^{<}\left( \varepsilon \right) &\smeq&\ci\left[\bm{ G}%
_{}^{r}\left( \varepsilon \right) \left(f_{L}(\varepsilon )\bm{ \Gamma }%
_{}^{L}\smpl f_{R}(\varepsilon )\bm{ \Gamma }_{}^{R}\right)\bm{ G}%
_{}^{a}\left( \varepsilon \right)\right]_{nm}  \nonumber \\
&\smeq&\bm{ G}_{nm}^{<}\left( \varepsilon \right) |_{V=0}-\ci\frac{eV}{2}\frac{%
\partial f(\varepsilon )}{\partial \varepsilon }\times \\
&&\left[ \bm{ G}^{r}\left( \varepsilon \right) (\bm{ \Gamma }^{L}\smmi\bm{ %
\Gamma }^{R})\bm{ G}^{a}\left( \varepsilon \right) \right] _{nm}\smpl{\cal O}%
(V^{2})
\nonumber
\label{linearized}
\end{eqnarray}
where $\bm{ G}^{r}(\varepsilon )\smeq[\varepsilon
\smmi \Ha\smmi\bm{\Sigma}_{L}^{r}\smmi\bm{\Sigma}_{R}^{r}]^{-1}$and $\bm{
G}^{a}\left(
\varepsilon \right) \smeq[\bm{ G}^{r}(\varepsilon )]^{\dagger }$ are the
retarded and advanced Green functions, respectively, $\bm{\Sigma}_{L}^{r}$ and $%
\bm{\Sigma}_{R}^{r}$ are the self-energies due to the left and right leads,
$f(\varepsilon )$ is the Fermi function and $\bm{ \Gamma }%
^{\eta }\smeq\ci(\bm{\Sigma}_{\eta }^{r}\smmi\bm{\Sigma}_{\eta }^{a})$ with $\eta \smeq
L,R$ .
The local spin polarization in the zero temperature limit and in the absence of an external magnetic
field, is then given by
\begin{equation}
\langle S_\mathrm{a}(\bm{ r}_{n})\rangle \smeq\frac{\hbar eV}{4}\mathrm{Tr}\left[
\sigma_\mathrm{a}\left\{\bm{ G}^{r}(\Ef) (\bm{ \Gamma }^{L}-\bm{ \Gamma
}^{R})\bm{ G}^{a}(\Ef)\right\} \right] _{nn}
\end{equation}
In what follows we use this expression to evaluate the CISP in different
geometries.
\begin{figure}[t]
   \centering
   \includegraphics[width=.37\textwidth]{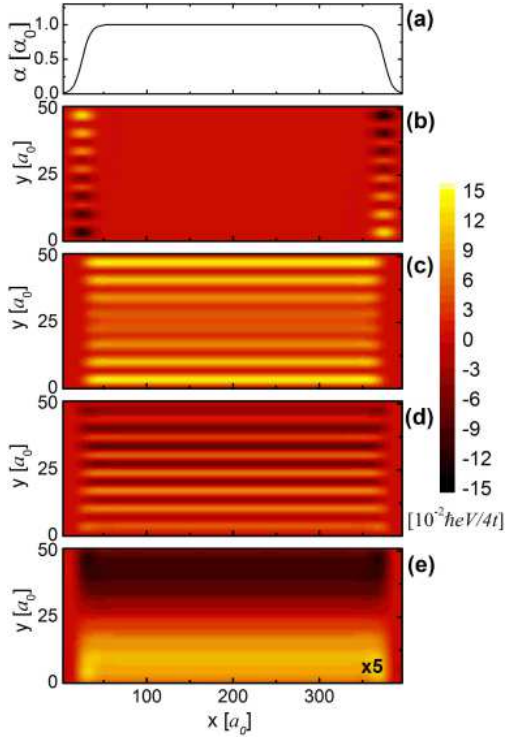}
   \caption{(color online) Current induced spin polarization in a quantum wire with
$L_x\smeq400a_0$, $L_y\smeq50a_0$, $\Ef\smeq5$meV and
$\alpha\smeq10$meVnm. (a) Spatial profile of the SO coupling
parameter $\alpha$; (b) $\langle S_{x}(x,y)\rangle$. Notice it is
very small inside the sample; (c) $\langle S_{y}(x,y)\rangle$; and
(d) $\langle S_{z}(x,y)\rangle$. Panel (e) shows the convolution
of $\langle S_{z}(x,y)\rangle$ with a Gaussian (we multiplied by a factor $5$ to keep
the same color scale).}
   \label{fig6}
\end{figure}
\subsection{Quantum wires}
We consider a two-terminal wire as shown in Fig. \ref{scheme}.
For the case of symmetric samples, as the one considerer here, all components of the
CISP have a well defined symmetry,
\begin{eqnarray}
\langle S_{x}(\bm{ R})\rangle_{V} &\smeq&-\langle S_{x}(-\bm{ R}%
)\rangle_{-V}  \nonumber \\
\langle S_{y}(\bm{ R})\rangle_{V} &\smeq&-\langle S_{y}(-\bm{ R}%
)\rangle_{-V} \nonumber \\
\langle S_{z}(\bm{ R})\rangle_{V} &\smeq&\langle S_{z}(-\bm{ R}%
)\rangle_{-V}
\end{eqnarray}
where $\bm{R}\smeq(x,y)$ is
the spatial coordinate measured from the sample's center. Moreover, a $\pi$-rotation of
the system along the $x$-axis gives
\begin{eqnarray}
\langle S_{x}(x,y)\rangle_{V,\alpha }
&\smeq&\langle S_{x}(x,-y)\rangle_{V,-\alpha
}\smeq-\langle S_{x}(x,-y)\rangle_{V,\alpha } \nonumber\\
\langle S_{y}(x,y)\rangle_{V,\alpha }
&\smeq&-\langle S_{y}(x,-y)\rangle_{V,-\alpha
}\smeq\langle S_{y}(x,-y)\rangle_{V,\alpha } \nonumber\\
\langle S_{z}(x,y)\rangle_{V,\alpha }
&\smeq&-\langle S_{z}(x,-y)\rangle_{V,-\alpha
}\smeq-\langle S_{z}(x,-y)\rangle_{V,\alpha }\nonumber\\
\end{eqnarray}
where the last equality is due to the symmetry of the Hamiltonian, $%
\Ha(\alpha ,\sigma _{x},\sigma _{y})\smeq \Ha(-\alpha ,-\sigma _{x},-\sigma _{y})$.
Similarly a $\pi$-rotation along the $y$-axis gives
\begin{eqnarray}
\langle S_{x}(x,y)\rangle _{V,\alpha }
&\smeq&-\langle S_{x}(-x,y)\rangle _{-V,-\alpha
}\smeq\langle S_{x}(-x,y)\rangle _{-V,\alpha }
\nonumber \\
\langle S_{y}(x,y)\rangle _{V,\alpha }
&\smeq&\langle S_{y}(-x,y)\rangle _{-V,-\alpha
}\smeq-\langle S_{y}(-x,y)\rangle _{-V,\alpha }\nonumber \\
\langle S_{z}(x,y)\rangle _{V,\alpha }
&\smeq&-\langle S_{z}(-x,y)\rangle _{-V,-\alpha
}\smeq-\langle S_{z}(-x,y)\rangle _{-V,\alpha }
\nonumber\\
\end{eqnarray}
In linear response, these relations imply that,\cite{YaoY05}
\bea
\langle S_{x}(x,y)\rangle &\smeq&-\langle S_{x}(-x,y)\rangle\smeq-\langle S_{x}(x,-y)\rangle\nonumber\\
\langle S_{y}(x,y)\rangle&\smeq&\langle S_{y}(-x,y)\rangle\smeq\langle S_{y}(x,-y)\rangle\nonumber\\
\langle S_{z}(x,y)\rangle&\smeq&\langle S_{z}(-x,y)\rangle\smeq-\langle
S_{z}(x,-y)\rangle\,.
\label{sim}
\eea
Note that in systems with translational invariance (or in the case of very large
samples), the first equation implies $\langle
S_{x}(x,y)\rangle\smeq0$, in agreement with the results of the
previous section. Any deviations from the symmetry relations shown in Eq. (\ref{sim}),
such as a larger spin polarization near the drain contact, can
only arise from non-linearities.\cite{NikolicSZS05}
\begin{figure}[t]
   \centering
   \includegraphics[width=.37\textwidth]{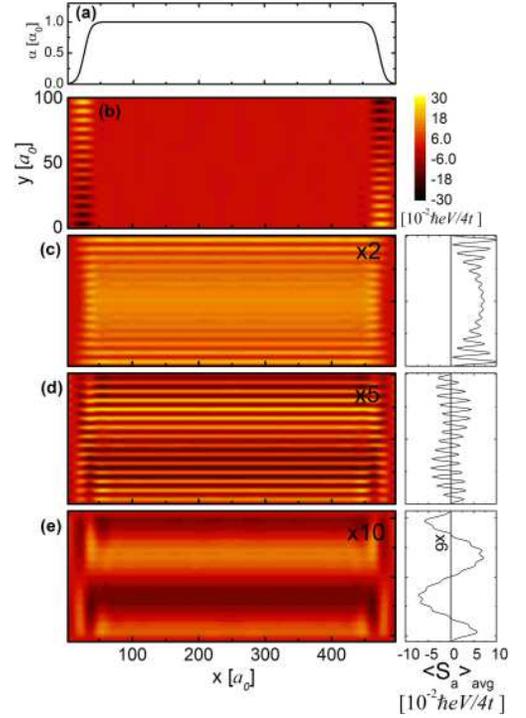}
   \caption{(color online) Same as in Fig. \ref{fig6} but for $L_x\smeq500a_0$, 
$L_y\smeq100a_0$ and $\Ef\smeq4.8$meV. The polarization densities were multiplied by the
factors indicated
in the figure to keep the same color scale).}
   \label{suave100}
\end{figure}

The transverse structure of the different components of $\langle \bm{ S}\rangle $
is determined by three characteristic lengths: the Fermi wavelength $\lambda_{F}$, the
SO length $L_{SO}\smeq\pi\hbar ^{2}/m^{*}\alpha \smeq a_{0}\pi t/\lambda $ and
the sample width $L_{y}$. For wide samples ($L_{y}/L _{SO}\!\gg\!1
$) it has been shown in the previous section that the CISP presents
transverse oscillations with the two characteristic lengths $\lambda _{F}$ and $L
_{SO}$. The resulting structure shows a beating of the two lengths with a
net spin polarization at the sample border.

Results for narrow samples ($L_{y}/L _{SO}\!\lesssim\!1$) are shown in Figs.
\ref{fig6} and \ref{suave100}. All the symmetries mentioned above are apparent in
the figures. The profile of the Rashba parameter $\alpha$ is shown in the top panels.
We notice that: (i) $\langle S_{x}\rangle$ (panel (b)) is
zero away from the sample-lead interfaces; (ii) $\langle S_{y}\rangle$ is the largest
spin component having a defined sign throughout the channel; and (iii) $\langle
S_{z}\rangle$ (panel (d)) oscillates with a dominant sign on each side of the
wire. To emphasize the later behavior, we made a convolution of $\langle
S_{z}\rangle $ with a Gaussian with an rms of three lattice parameters ($15nm$). The
convoluted spin density (panel (e)) clearly shows the spin polarization with
opposite signs at the two sides of the wire. As oppose to the wide sample case, however,
the transverse profile shows strong oscillations whose magnitude is comparable with the
value of the magnetization close to the edges.
\begin{figure}[t]
   \centering
   \includegraphics[width=.37\textwidth]{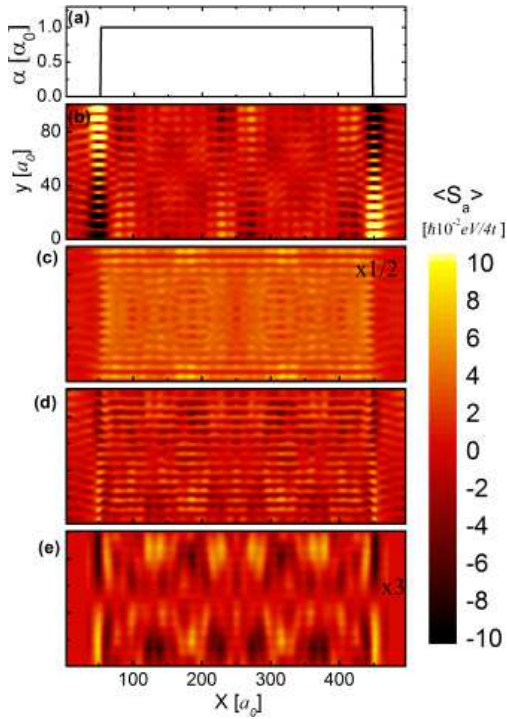}
   \caption{(color online) Same as in Fig. \ref{suave100} but for the case the SO
coupling $\alpha$ is
turned on abruptly at the sample-lead interfaces. Notice $\langle S_x\rangle\smneq0$
inside the sample and that the three components of the spin density show oscillations
along the $x$-axis.}
   \label{brusco100}
\end{figure}

\begin{figure}[b]
   \centering
   \includegraphics[width=.35\textwidth]{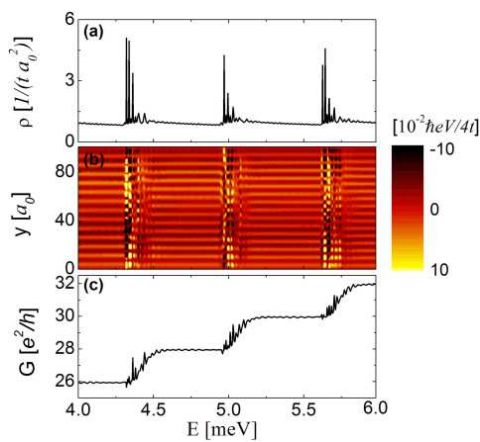}
   \caption{(color online) a) Local density of states for $x\smeq260a_0$; b) Energy
dependence of $\langle S_z(x=260a_0,y)\rangle$. Note that the number of oscillations in
the $y$-axis increases with the number of channels; and c) conductance as
a function of
$\Ef$. Parameters as in Fig. \ref{suave100}}
   \label{dwEf}
\end{figure}

It is interesting to note that the penetration of the $x$-component inside the sample
depends on the sample-lead interface. When the Rashba parameter $\alpha$ is turned on
abruptly, the interference induced by the now strong scattering at both interfaces
leads to oscillations of $\langle S_x\rangle$ along the $x$-axis---such oscillating
pattern is also
visible on the other two spin components. This is shown in Fig. \ref{brusco100}.
Another important feature is that in long samples ( $L_x/L_{SO}\!\gg\!1$), and when
$\alpha$ is turned one adiabatically, the finite size results away from the interface
are the same as those obtained in the previous section for infinite long systems.

The CISP is
dominated by the channel that contributes the most to the density of states (DOS) at the
Fermi level.  This is shown in Fig. \ref{dwEf}  where the
transverse structure of
$\langle S_{z}\rangle $ is plotted as a function of $\Ef$ for a fixed value of $x$.
Clearly the $z$-component of the
CISP is sensitive to the position of the Fermi
energy and the number of oscillations depends on the wavelength of the corresponding
transverse mode. In particular, if $\Ef$ lies close to the bottom of a channel,
the magnetization pattern is strongly modified: its magnitude increases and its sign can
change. 
This behavior suggest that when large voltages are applied to narrow samples, and
therefore a wide energy window around $\Ef$ is involved, non-linear effects might be
important due  to the contribution of states at the bottom of each channel. It also
shows the sensitivity of the CISP to the geometry while emphasizes the nature of the
effect: it arises from the interference of the different spin states which are mixed by
the scattering at the interface and the non-equilibrium occupancy of the momentum
states.

\begin{figure}[t]
   \centering
   \includegraphics[width=.4\textwidth]{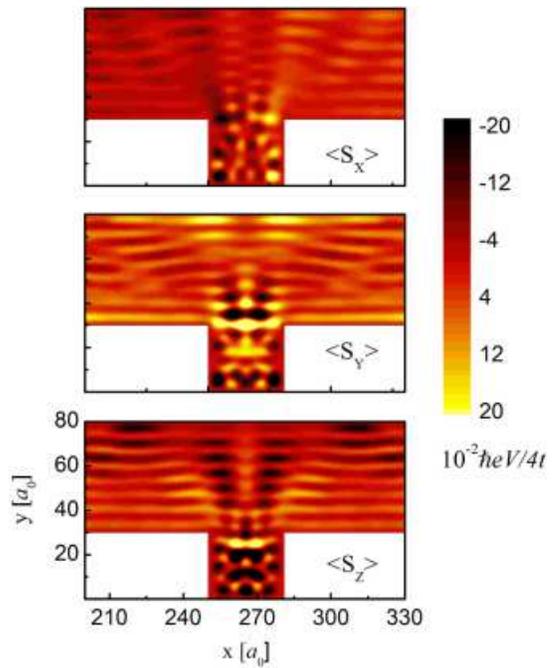}
   \caption{(color online) CISP in a `T' shaped sample. Electrons flows from left to
right. The left
and right lead are connected to the system at $x\smeq1$ and $x\smeq530$, respectively.}
   \label{fig11}
\end{figure}
\subsection{`T' and `L' shaped samples}

As mentioned in the previous sections, in the case of small samples,
both the magnitude and the profile of the CISP relies on the geometry of the system.
However, so far we have discussed only the wire geometry. In this section we will
discuss two different geometries: a `T' and a `L' shaped sample.

\begin{figure}[t]
   \centering
   \includegraphics[width=.4\textwidth]{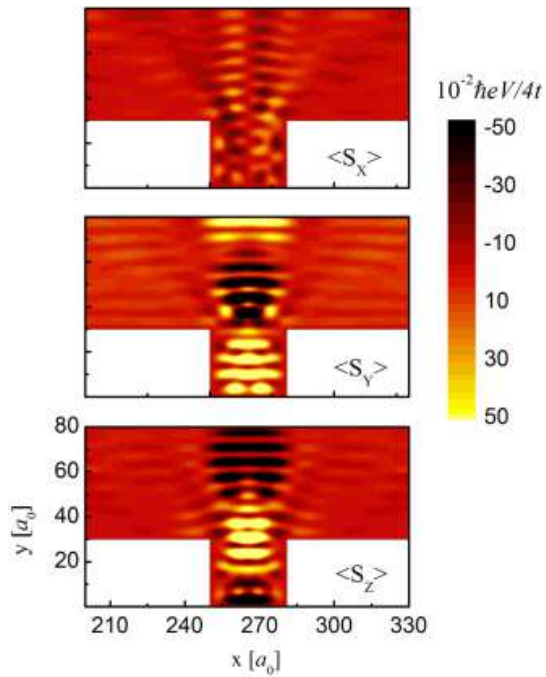}
   \caption{(color online) CISP in a `T' shaped sample when $\Ef$ is tuned to the value
of a resonant
state in the cavity.
   Notice the large amplification and localization of the induced polarization.}
   \label{fig12}
\end{figure}

Figures \ref{fig11} and \ref{fig12} show the three components of the CISP for the case
of a `T' shaped sample: it corresponds to a quantum wire with a open cavity attached to
it.
 The cavity is large enough to originate a (resonant) state which is
substantially localized around the cavity. Figure
\ref{fig11} correspond to an arbitrary value of $\Ef$ while in Fig. \ref{fig12} it is
tuned to correspond to a resonant state. Notice that all components are different from
zero. While in both cases there is some additional
polarization in the vicinity of the cavity, the latter case shows a strong enhancement.
We note that because the cavity is at the center of the system, some of the
symmetry relationships shown in Eq. (\ref{sim}) are still valid.
These results suggest that an adequate engineering of the sample geometry might
allow to design different profiles for the spin polarization, such as focusing the
spin polarization on a given spot, for instance.
Notably, the amount of spin polarization is very large---in systems without SO a
Zeeman field of $B\simeq1-10$T, would be required to achieve a similar
polarization density.
\begin{figure}[t]
   \centering
   \includegraphics[width=.3\textwidth]{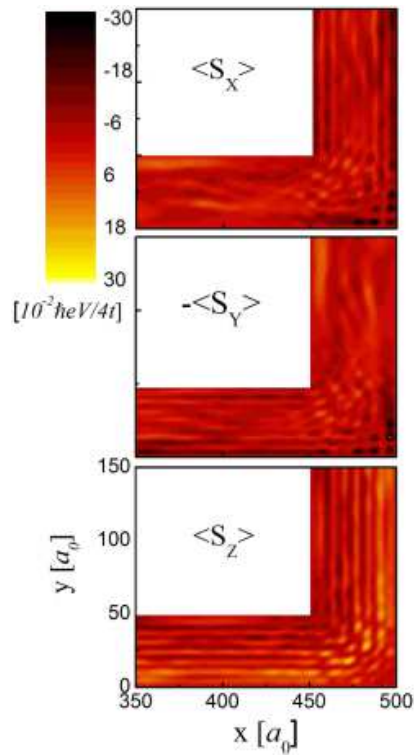}
   \caption{(color online) CISP in a L-shaped system. Notice the symmetry between the
transverse and longitudinal spin components.}
   \label{figL}
\end{figure}

Figure \ref{figL} shows the corresponding results for the  `L' shaped sample.
The presence of the `corner' induces
some additional spin polarization as compared to the wire geometry. The
symmetry between $\langle S_x\rangle$ and
$\langle S_y\rangle$ in the two `legs' of the `L' is apparent from the figure---to
emphasize this, we have plotted $-\langle S_y\rangle$ instead of  $\langle
S_y\rangle$. Because of the strong scattering at the corner, the longitudinal spin
component does not decay to zero, even far from it. This is similar to the case of an
abrupt turn on of the Rashba parameter.

\section{Summary and conclusions}
We have showed that current induced spin polarization is possible
in ballistic 2DEGs. The effect is purely geometrical and arises
from the electron scattering at the sample boundary. For this
reason, the CISP is strongly dependent on the shape of the sample
while some spin components can be very sensitive to the
sample-lead interface. In addition, the CISP is quite sensitive to
the value of the Fermi energy. This is true in both wide and
narrow samples, though for different reasons. In wide samples, the
CISP contains two distinct contribution: one comes from the
interference of the two components of the scattered wave, which are 
two bulk Bloch states from different bands, and the other from the evanescent modes that
appear at the boundary. Both contributions have different signs
and so tend to cancel each other. At low $\Ef$ the evanescent
modes clearly dominate so there is a large CISP. When the
value of $\Ef$ is increased, the cancellation between the two contributions is better
and the
CISP goes to zero. One interesting feature of the CISP for this
case is that it does not decay monotonically to zero as one goes from the boundary to
the bulk of the sample but oscillates. Then, even on a given side of the sample, the
polarization can change its sign. It is worth mentioning that because of this
strong decay with $\Ef$, the phenomena described here
\textit{cannot} account for the magnitude of the effect observed
in Ref. \onlinecite{SihMKLGA05}. In such device, the presence of a
strong disorder and an electric field seems to play an important role. One
possible explanation, within our intrinsic model, is that disorder
might rapidly destroy the phase of the scattered wave without affecting the evanescent
states too much, therefore unbalancing the cancellation of the two
contributions. In any case, we believe, that the special
properties of the scattering at the surface plays a crucial role
in originating the CISP in those samples.

The case of narrow systems is a bit different. There, the CISP is
dominated by the interference of the waves scattered at both sides of the
sample. Because of the strong confinement, the
density of states have some structure and then, depending on
$\Ef$, only a few states might dominate the CISP. This is more
clearly seen when $\Ef$ is close to the bottom of a transverse channel
or, in the presence of a side cavity, near a resonant state. In this
regime the profile of the CISP very much resembles that of a
single wavefunction. This opens up the possibility to design
different profiles for the CISP by modifying the geometry of the
sample.

\section{Acknowledgments}
This work was partially supported by ANPCyT
Grants No 13829 and 13476 and Fundaci\'on Antorchas, Grant 14169/21. AR and GU
acknowledge support from CONICET.


\begin{thebibliography}{56}
\bibitem[{not({\natexlab{a}})}]{note0}
\bibinfo{note}{We use the term 'polarization' instead of 'accumulation' to
  emphasize the fact that the effect discussed here does not result from the
  flow of a spin current that 'accumulate' spins at the boundary but rather
  from the difference in the occupation of the right-moving ($k_x>0$) and
  left-moving ($k_x<0$) states.}
\bibitem[{not({\natexlab{b}})}]{note2}
\bibinfo{note}{We only consider the case $\Ef>eV/2$.}
\end{thebibliography}
\end{document}